# Determination of the Third Neutrino-Mixing Angle $\theta_{13}$ and its Implications

## D. P. Roy

Homi Bhabha Centre for Science Education, Tata Institute of Fundamental Research, Mumbai 400088, India

**Abstract**

Till 2010 we had three unknown parameters of neutrino oscillation – the third mixing angle $\theta_{13}$, the sign of the larger mass difference $\Delta m_{31}^2$ and the CP violating phase $\delta$. Thanks to a number of consistent experimental results since then, culminating in the recent Daya Bay reactor neutrino data, we have a definitive determination of $\theta_{13}$ now. Moreover its measured value, $\sin^2 2\theta_{13} \approx 0.1$, is close to its earlier upper limit. This has promising implications for the determination of the two remaining unknown parameters from the present and proposed accelerator neutrino experiments in the foreseeable future. This article presents a pedagogical review of these profound developments for the wider community of young physicists including university students.

# 1. Introduction

Till 2010 our knowledge of the neutrino mass and mixing parameters was at the stage described by the fig. 1 and eqs. (1-3) below.

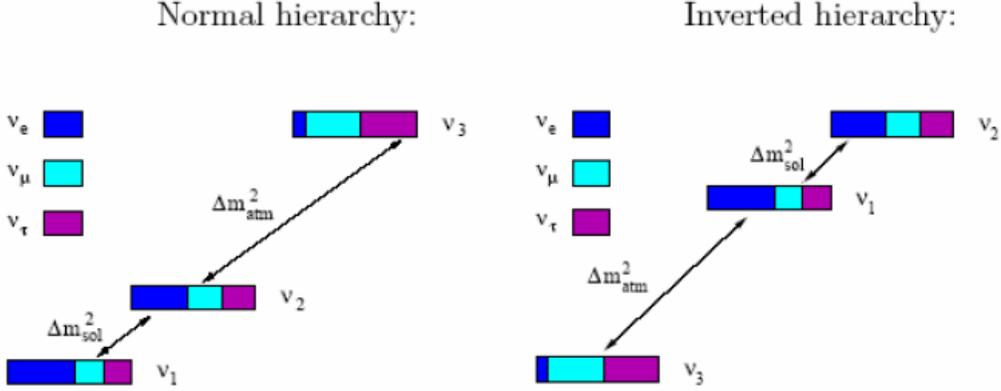

Fig 1. Schematic diagrams of the neutrino mass and mixing parameters [1].

The so called atmospheric and solar neutrino mass and mixing parameters had been measured to a few percent accuracy by various neutrino oscillation experiments [2], i.e.

$\Delta m_{atm}^2 = \Delta m_{32}^2 \approx \Delta m_{31}^2 \approx \pm 2.4 \times 10^{-3}$ eV$^2$, $\sin^2 2\theta_{23} \approx 1.0$; (1)

$\Delta m_{sol}^2 = \Delta m_{21}^2 \approx 7.6 \times 10^{-5}$ eV$^2$, $\sin^2 \theta_{12} \approx 0.3$. (2)

The magnitude of the larger mass square difference and the corresponding mixing angle of eq.(1) were determined by the atmospheric and long baseline (LBL) accelerator neutrino oscillation experiments, i.e. SK and MINOS/K2K respectively. The smaller mass square difference and the corresponding mixing angle of eq.(2) were determined by the solar neutrino experiments along with the LBL reactor neutrino experiment, KamLAND. Thanks to the solar matter effect, one could determine this mass square difference without the sign ambiguity and the mixing angle without the octant ambiguity. However, one had only an upper limit on the third mixing angle, representing the $\nu_e$ component of the 3$^{rd}$ mass eigenstate $\nu_3$, from the CHOOZ reactor neutrino experiment [3], i.e.

$\sin^2 2\theta_{13} < 0.15$ at 90% CL, (3)

for the $\Delta m_{atm}^2$ value of eq. (1). Moreover, the sign ambiguity of this parameter implied two alternative scenarios of normal and inverted mass hierarchy as indicated in fig. 1. Finally, the CP violating phase angle δ of the three neutrino mass matrix remained completely unknown. Thus the three primary goals of neutrino physics at this stage were the determination of (1) the size of the third mixing angle $\theta_{13}$, (2) the sign of the atmospheric mass difference $\Delta m_{atm}^2$ ($\Delta m_{31}^2$), and (3) the value of the CP violating phase $\delta$.

A series of indirect and direct neutrino oscillation experiments over the last two years have led to a consistent and by now definitive value of $\theta_{13}$. Let us start with a brief preview of these developments here, which will be discussed in the following sections with due reference to the sources. First came an indirect hint of a nonzero $\theta_{13}$ from a comparison of solar and KamLAND LBL reactor neutrino results in 2010. This was strengthened by the LBL accelerator neutrino experiment T2K in 2011, providing direct evidence of a nonzero $\theta_{13}$ at $2.5\sigma$ level. But the measured value of $\theta_{13}$ from this experiment was dependent on the other two unknowns, i.e. the sign of $\Delta m_{31}^2$ and the value of $\delta$. Finally came three short baseline (SBL) reactor neutrino experiments this year (2012) with increasing order of precision – i.e. Double Chooz, RENO and Daya Bay – giving a direct and definitive value of

$$\sin^2 2\theta_{13} \approx 0.1. \qquad (4)$$

Note that the measured value of $\theta_{13}$ is close to what was its upper limit (3) so far. This has profound implications for the determination of the other two unknown entities, the sign of $\Delta m_{31}^2$ via the earth matter effect and the value of the phase $\delta$, since the size of both the contributions are controlled by this angle. Thus the present time can be rightly called the watershed moment in the history of neutrino physics. This article presents a pedagogical review of these profound developments for a broad class of physicists including university students.

Section 2 gives a broad overview of the 3-neutrino mixing and oscillation formalism. This is followed by a brief description of the first indication of a nonzero $\theta_{13}$ from a comparison of solar and the KamLAND LBL reactor neutrino data in section 3. Then section 4 discusses the three above mentioned SBL reactor neutrino experiments, culminating in Daya Bay, which have provided the most direct and definitive determination of $\theta_{13}$. Section 5 discusses the LBL accelerator neutrino experiments, MINOS and T2K, along with the forthcoming NOvA experiment. Although the nonzero $\theta_{13}$ signal from T2K preceded those of the SBL reactor neutrino experiments, the resulting value of this angle was dependent on the unknown phase $\delta$ and the sign of $\Delta m_{31}^2$, as mentioned above. Indeed one hopes to use the future data from the T2K and NOvA experiments along with the precise knowledge of $\theta_{13}$ from the reactor neutrino data to get the first indications on the phase $\delta$ and the sign of $\Delta m_{31}^2$, as discussed in this section. We add a brief discussion of the role of future atmospheric neutrino experiments in determining the sign of $\Delta m_{31}^2$ in section 6. We conclude with a summary of the main points in section 7.

## 2. Three Neutrino Mixing and Oscillation Formalism

The three neutrino flavour eigenstates are related to the three mass eigenstates through the formula

$$\nu_\alpha = \sum U^*_{\alpha i} \nu_i, \alpha = e, \mu, \tau \tag{5}$$

where the mixing matrix U is a 3x3 unitary matrix described by the three above mentioned mixing angles and the phase δ. This is analogous to the CKM matrix for the quark sector and called PMNS matrix after the authors of neutrino mixing and oscillation [4]. It can be written either in a compact form or expanded as a product of three 2x2 rotation matrices for better insight, i.e.

$$U = \begin{bmatrix} c_{12}c_{13} & s_{12}c_{13} & s_{13}e^{-i\delta} \\ -s_{12}c_{23} - c_{12}s_{23}s_{13}e^{i\delta} & c_{12}c_{23} - s_{12}s_{23}s_{13}e^{i\delta} & s_{23}c_{13} \\ s_{12}s_{23} - c_{12}c_{23}s_{13}e^{i\delta} & -c_{12}s_{23} - s_{12}c_{23}s_{13}e^{i\delta} & c_{23}c_{13} \end{bmatrix}$$

$$= \begin{bmatrix} 1 & 0 & 0 \\ 0 & c_{23} & s_{23} \\ 0 & -s_{23} & c_{23} \end{bmatrix} \begin{bmatrix} c_{13} & 0 & s_{13}e^{-i\delta} \\ 0 & 1 & 0 \\ -s_{13}e^{i\delta} & 0 & c_{13} \end{bmatrix} \begin{bmatrix} c_{12} & s_{12} & 0 \\ -s_{12} & c_{12} & 0 \\ 0 & 0 & 1 \end{bmatrix}$$

$$\tag{6}$$

where $c_{ij}$ and $s_{ij}$ denote $\cos\theta_{ij}$ and $\sin\theta_{ij}$ respectively [2,5,6,7]. Note that the three mixing angles are simply related to the flavour components of the three mass eigenstates as

$$\frac{|U_{e2}|^2}{|U_{e1}|^2} = \tan^2\theta_{12}, \frac{|U_{\mu 3}|^2}{|U_{\tau 3}|^2} = \tan^2\theta_{23}, |U_{e3}|^2 = \sin^2\theta_{13}. \tag{7}$$

The vacuum oscillation probability between two neutrino flavours α and β in the three neutrino oscillation formalism is given by

$$P(\nu_\alpha \to \nu_\beta) = \left| \sum_j U_{\beta j} e^{\frac{-im_j^2 L}{2E_\nu}} U^*_{\alpha j} \right|^2, \tag{8}$$

where the last factor comes from the decomposition of $\nu_\alpha$ into the mass eigenstates, the phase factor in the middle from the propagation of each mass eigenstate over distance L, and the first factor from their recomposition into the flavour eigenstate $\nu_\beta$ at the end. Proceeding exactly as in the case of two flavour oscillation, one can rewrite the above formula as [6,7]

$$P(\nu_\alpha \to \nu_\beta) = \delta_{\alpha\beta} - 4\sum_{i>j} \text{Re}\left[U^*_{\alpha i} U_{\alpha j} U_{\beta i} U^*_{\beta j}\right] \sin^2 \Delta_{ij}$$
$$+ 2\sum_{i>j} \text{Im}\left[U^*_{\alpha i} U_{\alpha j} U_{\beta i} U^*_{\beta j}\right] \sin 2\Delta_{ij}, \tag{9}$$

where

$$\Delta_{ij} = \Delta m^2_{ij} L / 4 E_\nu. \tag{10}$$

The last term of eq. (9) contains the CP violating contribution, proportional to sin δ. Note that this contribution can only be measured in neutrino oscillation experiments measuring the appearance probability of a new flavour, since for disappearance experiments (β = α) the last term vanishes identically. Moreover, the CP violating contribution changes sign in going from $P(\nu_\alpha \to \nu_\beta)$ to $P(\nu_\beta \to \nu_\alpha)$. It also changes sign in going form $P(\nu_\alpha \to \nu_\beta)$ to $P(\bar\nu_\alpha \to \bar\nu_\beta)$ since $P(\nu_\beta \to \nu_\alpha) = P(\bar\nu_\alpha \to \bar\nu_\beta)$ by CPT invariance.

Using the identity

$$\Delta m^2_{32} = m^2_3 - m^2_2 \Rightarrow \Delta_{32} = \Delta_{31} - \Delta_{21} \tag{11}$$

one can express the vacuum oscillation probability (9) in terms of sinusoidal functions of the two independent mass scales on the RHS [5-7]. Moreover, one can use the observed hierarchy between the two mass scales,

$$\alpha = \left|\Delta m^2_{21}\right| / \left|\Delta m^2_{31}\right| = \left|\Delta_{21}\right| / \left|\Delta_{31}\right| \cong 0.03, \tag{12}$$

to write this probability in terms of a single mass scale to a very good approximation. For this purpose it is useful to rewrite $\Delta_{ij}$ (10) in terms of convenient units, i.e.

$$\Delta_{ij} = 1.27 \Delta m^2_{ij} L / E_\nu, \tag{13}$$

where $\Delta m_{ij}^2$ is in eV$^2$, distance L is in km(m) and the neutrino energy $E_\nu$ is in GeV(MeV). Thus we see from (1,2) that for atmospheric or LBL accelerator neutrino experiments

$$E_\nu \approx GeV, L \approx 10^3 km \Rightarrow \Delta_{31} \approx 1, \Delta_{21} \approx \alpha, \tag{14}$$

so that the dominant contribution to the oscillation probability (9) comes from the $\Delta_{31}$ scale. The same is true for the SBL reactor neutrino experiments, where $E_\nu \approx$ MeV and L $\approx 10^3$ m. Thus to a very good approximation we have the $\nu_\mu$ survival probability [5]

$$P(\nu_\mu \to \nu_\mu) \cong 1 - (c_{13}^4 \sin^2 2\theta_{23} + s_{23}^2 \sin^2 2\theta_{13})\sin^2\Delta_{31}$$
$$\cong 1 - \sin^2 2\theta_{23} \sin^2 \Delta_{31}, \tag{15}$$

where we have neglected terms of order $\sin^2\theta_{13}\cos 2\theta_{23}$ and $\sin^4\theta_{13}$ in the final step following eqs. (1) and (3). Thus the expression reduces to the simple two neutrino mixing formula to a very good approximation. This means that the values of $\Delta m_{31}^2$ and $\sin^2 2\theta_{13}$ obtained from atmospheric and LBL accelerator neutrino experiments using this simple formula hold good to a very high degree of accuracy. On the other hand, it also means that these experiments are not very useful for the determination of $\theta_{13}$ because of their insensitivity to this parameter. The corresponding expression for the $\nu_e$ survival probability is

$$P(\nu_e \to \nu_e) \cong 1 - \sin^2 2\theta_{13} \sin^2 \Delta_{31}, \tag{16}$$

and an identical expression for the corresponding $\bar{\nu}_e$ survival probability, which is used in determining $\sin^2 2\theta_{13}$ from SBL reactor neutrino experiments discussed in section 4. Note that for the KamLAND LBL reactor neutrino experiment we have

$$E_\nu \approx MeV, L \approx 10^5 m \Rightarrow \Delta_{31} \approx 1/\alpha, \Delta_{21} \approx 1, \tag{17}$$

so that the oscillation terms in $\Delta_{31}$ approach their average values over a complete cycle. Thus the vacuum $\nu_e$ survival probability is again given in terms of a single scale to a good approximation [5], i.e.

$$P(\nu_e \to \nu_e) \cong 1 - \tfrac{1}{2}\sin^2 2\theta_{13} - c_{13}^4 \sin^2 2\theta_{12} \sin^2 \Delta_{21}$$
$$\cong c_{13}^4(1 - \sin^2 2\theta_{12} \sin^2 \Delta_{21}), \tag{18}$$

neglecting the $s_{13}^4$ terms in the last step. This formula is used in estimating $\theta_{13}$ from a comparison of solar and KamLAND reactor neutrino experiments, discussed in section 3.

For the appearance probability $P(\nu_\mu \to \nu_e)$, the leading scale ($\Delta_{31}$) contribution is suppressed by a small coefficient $\sim \sin^2 2\theta_{13}$. Therefore one has to consider the subleading scale contributions as well. The full expression for this vacuum oscillation probability is given by [5]

$$P(\nu_\mu \to \nu_e) = \sin^2 2\theta_{13} s_{23}^2 \sin^2 \Delta_{31}$$
$$+ \alpha \sin 2\theta_{13} \sin 2\theta_{12} \sin 2\theta_{23} \cos(\Delta_{31}+\delta)\Delta_{31}\sin\Delta_{31} + \alpha^2 \sin^2 2\theta_{12} c_{23}^2 \Delta_{31}^2. \tag{19}$$

Here the first term representing the leading (atmospheric) scale contribution is suppressed by $\sin^2 2\theta_{13}$, while the last term representing the subleading (solar) scale contribution is suppressed by $\alpha^2$. The second term represents the CP violating and CP conserving parts of the interference term, which are suppressed by $\alpha\sin2\theta_{13}$. Of course we know now from eqs. (4) and (12) that $\sin2\theta_{13} \sim 1/3$, while $\alpha \sim 1/30$. Thus the interference term and in particular the CP violating contribution is suppressed by a factor of $\sim 10$, while the last term is suppressed by a factor of $\sim 100$ relative to the first term. Finally, note that the corresponding expression for $P(\nu_e \to \nu_\mu)$ or $P(\bar{\nu}_\mu \to \bar{\nu}_e)$ is obtained from (19) by simply changing the sign of the phase $\delta$.

Finally let us consider the earth matter effect on the above $\nu_e$ appearance probability, which will be relevant for the discussion of the LBL accelerator neutrino experiments in section 5. It comes from the charged current interaction of $\nu_e$ with electrons resulting in a potential energy term

$$V = \sqrt{2}G_F N_e \cong 7.6\times10^{-14}\left(\frac{\rho}{g/cm^3}\right)Y_e \text{ eV}, \tag{20}$$

where $G_F$ is the Fermi coupling and $N_e$ the electron number density in the terrestrial matter. For $\nu_e$ passing through the earth's crust one can write this in terms of a nearly constant matter density and electron fraction per nucleon,

$$\rho \cong 3g/cm^3, Y_e \cong 0.5. \tag{21}$$

In order to calculate the neutrino oscillation probability in matter one has to solve the Schrodinger equation for the neutrino state vector in the flavor basis,

$$i\frac{d}{dt}|\nu(t)\rangle = H|\nu(t)\rangle, \tag{22}$$

with the effective Hamiltonian

$$H \approx \frac{1}{2E_\nu}U diag(0,\Delta m^2_{21},\Delta m^2_{31})U^\dagger + diag(V,0,0). \tag{23}$$

For antineutrinos one has to make the replacements

$$U \to U^*, V \to -V. \tag{24}$$

For the case of constant matter density one can diagonalize the effective Hamiltonian perturbatively giving

$$H = U' diag(E_1,E_2,E_3)U'^\dagger, \tag{25}$$

where the expressions for the eigenvalues $E_{1,2,3}$ and the mixing matrix in matter U' can be found in ref [8]. Then the resulting oscillation probability is given in terms these quantities, i.e.

$$P(\nu_\alpha \to \nu_\beta) = \left| \sum_j U'_{\beta j} e^{-iE_j L} U'^*_{\alpha j} \right|^2,  \qquad (26)$$

which is analogous to the vacuum oscillation formula (8).

One can expand the oscillation probability in powers of $s_{13}$ and $\alpha$. Up to second order terms in $s_{13}$ and $\alpha$ one gets a fairly simple analytical formula [8]

$$P(\nu_\mu \to \nu_e) = 4 s_{13}^2 s_{23}^2 \frac{\sin^2(A-1)\Delta_{31}}{(A-1)^2} + \alpha^2 \sin^2 2\theta_{12} c_{23}^2 \frac{\sin^2 A\Delta_{31}}{A^2}$$
$$+ 2\alpha s_{13} \sin 2\theta_{12} \sin 2\theta_{23} \cos(\Delta_{31} + \delta) \frac{\sin A\Delta_{31}}{A} \frac{\sin(A-1)\Delta_{31}}{A-1},  \qquad (27)$$

where

$$A = \frac{VL}{2\Delta_{31}} = \frac{2E_\nu V}{\Delta m_{31}^2} \cong \pm \frac{E_\nu (GeV)}{10}.  \qquad (28)$$

This is analogous to the vacuum oscillation formula (19) to which it reduces for $A \to 0$. The matter effect is represented by the dimensionless quantity A. Note that the sign of A changes with the sign of $\Delta m_{31}^2$ as well as in going from neutrino to the corresponding antineutrino experiment. The former implies that the matter effect can be used to determine the sign of $\Delta m_{31}^2$, while the latter implies that it can fake a CP violating effect and hence complicate the extraction of $\delta$ by comparing neutrino and antineutrino data. For off-axis experiments like T2K and NOvA discussed in section 5, the typical beam energy is $E_\nu \sim 1$ GeV, so that one can expand (27) in powers of A [9]. Keeping only terms up to the first order in A we get

$$P(\nu_\mu \to \nu_e) = 4 s_{13}^2 s_{23}^2 [\sin^2 \Delta_{31} + A(2\sin^2 \Delta_{31} - \Delta_{31} \sin 2\Delta_{31})] + \alpha^2 \sin^2 2\theta_{12} c_{23}^2 \Delta_{31}^2$$
$$+ 2\alpha s_{13} \sin 2\theta_{12} \sin 2\theta_{23} \cos(\Delta_{31} + \delta) \Delta_{31}[\sin \Delta_{31} + A(\sin \Delta_{31} - \Delta_{31} \cos \Delta_{31})].$$
$$(29)$$

For optimal $\nu_\mu \to \nu_e$ appearance experiments $\Delta_{31} \sim \pi/2$, so that $\cos \Delta_{31}$ and $\sin 2\Delta_{31} \sim 0$. Thus the relative size of the matter effect in the leading term is $\sim 2A$.

## 3. First Hint of a Nonzero $\theta_{13}$

The first hint of a nonzero $\theta_{13}$ came from a joint analysis of the KamLAND LBL reactor (anti)neutrino along with the global solar neutrino data. As we saw from eq. (18), the vacuum survival probability of $\nu_e$ in the 3-neutrino mixing formalism is given by $\cos^4\theta_{13}$ times that of the 2-neutrino mixing case,

$$P_3(\nu_e \to \nu_e) \cong c_{13}^4 P_2(\nu_e \to \nu_e). \tag{30}$$

Essentially the same relation holds even in the presence of the solar matter effect [10]. Let us recall that in the 2-netrino case the solar matter effect is described by the MSW formalism [11] inside a triangular region of the $\Delta m_{21}^2$ - $\sin^2\theta_{12}$ parameter space, defined by the level-crossing and the adiabatic conditions [1]

$$\frac{\Delta m_{21}^2 \cos 2\theta_{12}}{2\sqrt{2} G_F N_e^0} < E_\nu < \frac{\Delta m_{21}^2 \sin^2 2\theta_{12}}{2\cos 2\theta_{12} (dN_e/dlN_e)_C}, \tag{31}$$

where $N_e^0$ is the electron density at the solar core and the last factor in the parenthesis is the fractional electron density gradient in the level-crossing region of the two neutrino energy eigenvalues. Inside this triangular region the $\nu_e$ produced at the solar core emerges as the higher mass eigenstate $\nu_2$ after passing through the level-crossing region. The resulting $\nu_e$ survival probability is given by the square of the $\nu_e$ component of $\nu_2$, i.e. $\sin^2\theta_{12}$. In the 3-neutrino oscillation formalism the electron density $N_e$ in (31) should be replaced by $N_e c_{13}^2$, resulting in a marginal shift of the boundary [12]. However, the solar neutrino mixing angle is primarily determined by the SK and SNO experiments with $E_\nu$ values deep inside this triangular region [1,2]. Thus the only change in going to the 3-neutrino oscillation formalism comes from the above mentioned renormalization factor of $c_{13}^4$, i.e.

$$P_{solar}(\nu_e \to \nu_e) \cong c_{13}^4 \sin^2\theta_{12}. \tag{32}$$

One sees from eq. (18) that for a nonzero $\theta_{13}$ the renormalization factor of $c_{13}^2$ would imply that the vacuum $\nu_e$ survival probability measured from the KamLAND data in the 2-neutrino oscillation formalism gives a slightly larger $\theta_{12}$ than its real value. On the other hand, eq. (32) implies that the corresponding measurement from the solar neutrino data would give a slightly smaller $\theta_{12}$ than its real value. Thus a slightly larger value of $\theta_{12}$ from the KamLAND data relative to that obtained from the solar neutrino data in the 2-neutrino oscillation formalism would indicate a nonzero $\theta_{13}$ [13]. An improved analysis of the SNO solar neutrino data [14], with a lower $E_\nu$ threshold and lower systematic errors, showed the central value of the solar $\theta_{12}$ to be slightly smaller than that of KamLAND in the 2-neutrino oscillation formalism, as expected for a nonzero $\theta_{13}$. A joint analysis of both these data in a 3-neutrino oscillation formalism led to

$$s_{13}^2 = \sin^2\theta_{13} = 2.0^{+2.1}_{-1.6} \times 10^{-2}, \tag{33}$$

giving the first hint of a nonzero $\theta_{13}$ at 1σ level. Combining this with all other data, including the preliminary $\nu_\mu \to \nu_e$ appearance data from the MINOS experiment, improved this to 2σ level [15], with

$$\sin^2\theta_{13} = 2 \pm 1 \times 10^{-2}. \tag{34}$$

Still this was no more than an experimental hint of a nonzero $\theta_{13}$.

## 4. Determination of $\theta_{13}$ by SBL Reactor (Anti)neutrino Expts.

The unambiguous and by now fairly precise determination of $\theta_{13}$ has come this year (2012) from three reactor (anti)neutrino experiments described below in increasing order of precision – i.e. Double Chooz [16, 17], RENO [18] and Daya Bay [19,20].

**Double Chooz Experiment:** The detector consists of a cylindrical target containing 10 m$^3$ of Gadolinium doped liquid scintillator to detect the reactor antineutrino via its inverse beta decay process

$$\bar{\nu}_e + p \to e^+ + n, \tag{35}$$

by recording the prompt signal of positron in the scintillator, followed by that of a ~ 8 MeV γ-ray coming from the neutron capture in Gd. The target is surrounded by a 55 cm thick concentric cylinder of undoped liquid scintillator (γ catcher) to detect the γ-rays escaping from the edge of the target cylinder. The system is surrounded in turn by 390 photo-multiplier tubes (PMT) to measure the scintillation energy. The synchronization between the positron and the γ detections helps to reduce the background to ~ 10% of the signal size. The detector is placed at a distance L = 1050 m from the 2x4.25 GW Chooz reactor complex in France. The experiment plans to shortly add a nearby detector of similar material to reduce the systematic error arising from the estimations of $\bar{\nu}_e$ flux and detector efficiency. In the absence of the near detector the flux is estimated from the reactor power, resulting in a fairly large systematic error. The first result from this experiment after 101 live days run [16] reported 4121 events against the no oscillation ($\theta_{13} = 0$) prediction of 4344 ± 165 events. The ratio

R = 0.944 ± 0.016 (stat) ± 0.040 (syst), (36)

corresponding to the $\nu_e$ survival probability (16), provides a 1.7σ evidence for nonzero $\theta_{13}$. The prompt positron energy measured by the scintillator (including its annihilation energy with an electron in the detector)

$E_{prompt} = E_\nu + m_p - m_n + m_e \approx E_\nu - 0.8$ MeV, (37)

was also found to show a spectral distortion as expected from the oscillation formulae (16) and (13). Combining the two results gave

$$\sin^2 2\theta_{13} = 0.086 \pm 0.041 \text{ (stat)} \pm 0.034 \text{ (syst)}. \tag{38}$$

Subsequently the experiment has reported the result of 228 days run [17], which doubles the statistics to ~ 8000 events. A combined analysis of rate and spectral distortion of these events gives

$$\sin^2 2\theta_{13} = 0.109 \pm 0.030 \text{ (stat)} \pm 0.025 \text{ (syst)}, \tag{39}$$

which provides a ~ 3σ signal for nonzero $\theta_{13}$.

**RENO Experiment:** It detects the antineutrino coming from an array of 6x2.8 GW reactors at the Yongwang Nuclear Power Plant in Korea, which are roughly equispaced on a line spanning ~ 1.3 km. It uses two identical detectors placed on the perpendicular bisector of the reactor array, at distances of 294 m (near) and 1383 m (far) from the array centre. Each detector consists of a cylindrical target containing 16 tons (18.6 m$^3$) of Gd-doped liquid scintillator to detect the prompt positron coming from the inverse beta decay process (35) along with the delayed γ-ray coming from the neutron capture in Gd. This is surrounded by a 60 cm thick concentric cylinder of undoped liquid scintillator (γ catcher) which is surrounded in turn by 354 PMT to measure the scintillation energy. The synchronization between the detections of the positron and the γ-ray reduces the background to ~ 3% (6%) of the signal in the near (far) detector. The RENO collaboration has reported observation of 17102 (154088) $\bar{\nu}_e$ events in the far (near) detector based on 229 days data [18]. In the absence of neutrino oscillation ($\theta_{13} = 0$) one can predict the no of signal events in the far detector relative to those in the near detector by rescaling the latter by a weighted average of the relative flux factors $(L_i^n/L_i^f)^2$ over the 6 reactors times the relative detection efficiency factor ($\varepsilon^f/\varepsilon^n$). They find a clear deficit of ~ 8% in the number of observed events in the far detector relative to this prediction, i.e.

$$R = 0.920 \pm 0.009 \text{ (stat)} \pm 0.014 \text{ (syst)}. \tag{40}$$

Fitting this deficit factor to the spectrum averaged oscillation formulae (16) and (13) gives

$$\sin^2 2\theta_{13} = 0.113 \pm 0.013 \text{ (stat)} \pm 0.019 \text{ (syst)}, \tag{41}$$

which constitutes a 4.9σ signal for a nonzero $\theta_{13}$. Moreover, the measured prompt energy distribution (37) shows evidence of spectral distortion as expected from these oscillation formulae.

**Daya Bay Experiment:**

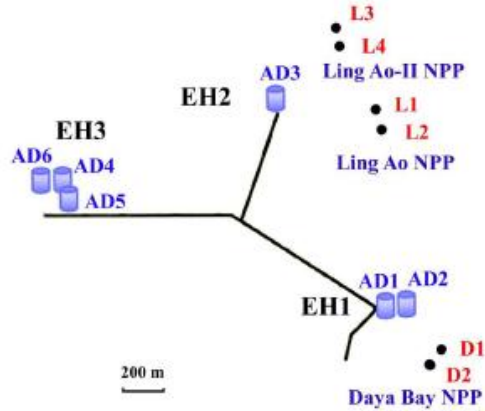

Fig 2. Layout of the Daya Bay experiment at the Daya Bay and Ling Ao Nuclear Power Plant (NPP) complex in China [19]. The dots represent the locations of the six reactors. Three of the six detectors are placed in the experimental halls EH1 and EH2 near the reactors, while the remaining three are placed in the far experimental hall EH3.

This is the most powerful of the three SBL reactor (anti)neutrino experiments, detecting the $\bar{\nu}_e$ coming from 6x2.9 GW reactors in 6 identical detectors – 3 near and 3 far from the reactor complex. It is also the most complex one in terms of the reactor and detector layouts, as shown in fig. 2 [19]. The flux-weighted baseline lengths of the two near detector halls are 470 m and 576 m, while that of the far experimental hall is 1648 m. Each detector consists of a cylindrical target containing 20 tons of Gd-doped liquid scintillator, surrounded by a concentric cylinder containing 20 tons of undoped liquid scintillator (γ catcher). The latter is surrounded by 192 PMT to measure the scintillation energy. Synchronization between the detections of the prompt positron coming from the inverse beta decay process (35) and the delayed γ-ray coming from the neutron capture in Gd reduces the background to ~ 2% (5%) of the signal in the near (far) detectors. The first result from this experiment based on only 55 days data reported observation of 10416 (80376) $\bar{\nu}_e$ events in the far (near) detectors [19]. In the absence of neutrino oscillation ($\theta_{13}$ = 0) one can again predict the number of $\bar{\nu}_e$ signal events in the far detectors (EH3) relative to those in the near ones (EH1 and EH2) following the above mentioned prescription. Here the baseline length of the near detectors $L_i^n$ corresponds to the flux-weighted average of those in EH1 and EH2 with respect to the ith reactor. There was a clear deficit of 6% in the number of observed signal events in the far detectors relative to this prediction, i.e.

R = 0.94 ± 0.011 (stat) ± 0.004 (syst). (42)

Fitting this deficit factor with the spectrum averaged oscillation formulae (16) and (13) gives a value of

$\sin^2 2\theta_{13} = 0.092 \pm 0.016$ (stat) $\pm 0.005$ (syst), (43)

which constitutes a 5.2σ signal for a nonzero $\theta_{13}$. The observed distribution of the prompt energy (37) also shows the expected spectral distortion from these oscillation formulae. Recently the Daya Bay collaboration has presented the result of their 140 days data [20] showing a deficit factor of

$R = 0.944 \pm 0.007$ (stat) $\pm 0.003$ (syst). (44)

It corresponds to an impressive 7.7σ signal for

$\sin^2 2\theta_{13} = 0.089 \pm 0.010$ (stat) $\pm 0.005$ (syst). (45)

The low systematic error of this experiment has been attributed mainly to ensuring the identity of the detectors from the beginning of their fabrication [21]. They plan to add one near and one far detectors in the halls EH1 and EH3 this year, and project a ~ 5% precision in the $\sin^2 2\theta_{13}$ measurement in 3 years [21].

A global summary of all the signals of nonzero $\sin^2 2\theta_{13}$ is shown in fig.3 [21]. It includes the signals from the LBL accelerator neutrino experiments of MINOS and T2K to be discussed in the next section. In particular the MINOS result is seen to be very sensitive to the choice of the mass hierarchy (sign of $\Delta m_{31}^2$). They are also sensitive to value of the CP violating phase δ (shown here for δ = 0). The effect of the recent reevaluation of the reactor $\bar{\nu}_e$ flux [22] on the KamLAND and the first Double Chooz results are also indicated. It does not affect the RENO and Daya Bay results, since they use the near detectors to measure the reactor $\bar{\nu}_e$ flux. One sees a consistent picture of a nonzero $\theta_{13}$ emerging over the past two years, with the most precise value given by the Daya Bay update of eq. (45). This is indicated by the narrow band. On the other hand, taking a weighted average of the three final results of Double Chooz, RENO and Daya Bay experiments gives

$\sin^2 2\theta_{13} = 0.097 \pm 0.010$, (46)

where all the errors have been added in quadrature. The central value corresponds to $\theta_{13} \approx 9°$.

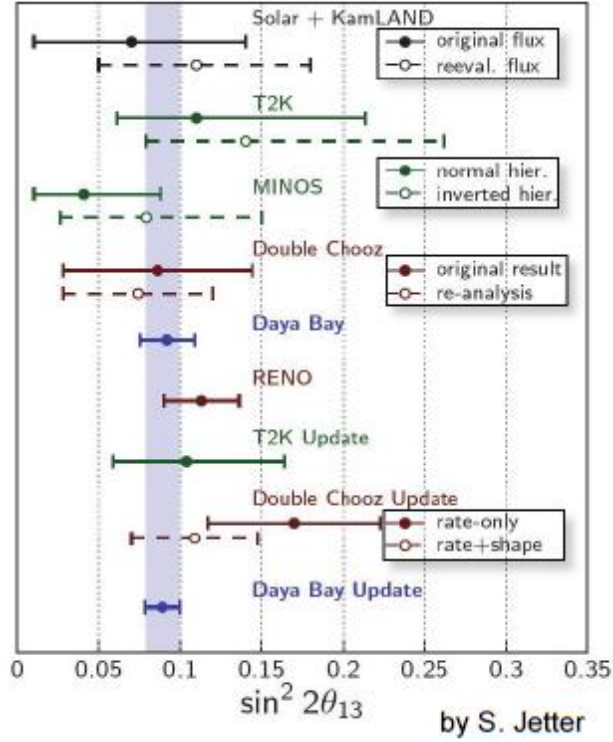

Fig 3. Global summary of the consistent evolution of a nonzero $\theta_{13}$ signal, culminating in the latest Daya Bay result [21].

## 5. Implications for LBL Accelerator Neutrino Experiments

As mentioned earlier, the $\nu_\mu \to \nu_e$ appearance data from the LBL accelerator neutrino experiments of MINOS and especially T2K had provided evidence for a nonzero $\theta_{13}$ ahead of the reactor neutrino data. However, the resulting value of $\theta_{13}$ was dependent on the remaining two unknown parameters - the sign of $\Delta m_{31}^2$ and the value of the CP violating phase $\delta$. Indeed the chief merit of these accelerator neutrino measurements lie in their sensitivity to these two unknown parameters, because it offers the possibility of determining them from the $\nu_\mu \to \nu_e$ appearance data of the present and proposed accelerator neutrino experiments using the precise value of $\theta_{13}$ from the forthcoming reactor neutrino data as input. Before discussing these experiments, however, a brief discussion of the accelerator neutrino beam is in order.

**On-axis and Off-axis Experiments:** The accelerator neutrino beam originates from the collision of the extracted proton beam on a solid target like aluminum or graphite, producing $\pi$ mesons. The resulting $\pi^+$ mesons are magnetically focused along the proton beam axis. Then their main decay process,

$$\pi^+ \to \mu^+ \nu_\mu, \tag{47}$$

produces the desired neutrino beam. It follows from the standard 2-body decay kinematics that the energy of the neutrino emerging at a small angle θ relative to the beam axis is related to the pion energy via

$$E_\nu \cong (1 - \frac{m_\mu^2}{m_\pi^2}) \frac{E_\pi}{1 + \gamma^2 \theta^2} \cong \frac{1}{2} \frac{E_\pi}{1 + \gamma^2 \theta^2}, \gamma = E_\pi / m_\pi. \qquad (48)$$

The corresponding neutrino flux per unit area of a detector placed at a distance r from the $\pi^+$ decay point is given by [7]

$$\Phi = \left(\frac{2\gamma}{1 + \gamma^2 \theta^2}\right)^2 \frac{1}{4\pi r^2}. \qquad (49)$$

We see from (48) and (49) that one gets the largest neutrino energy and flux for on-axis (θ = 0) neutrino experiments. All the first generation accelerator neutrino experiments including K2K and MINOS were on-axis experiments. However, these experiments are not well suited for $\nu_\mu \to \nu_e$ appearance because of two serious backgrounds. Firstly, the on-axis neutrino beam has a large $E_\nu$ (≈ $E_\pi$ /2) tail from that of $E_\pi$. It results in a serious neutral current background from

$$\nu_\mu p \to \nu_\mu p \pi^0, \ \pi^0 \to \gamma\gamma, \qquad (50)$$

followed by the $\gamma \to e^+ e^-$ pair creation. Secondly, the on-axis $\nu_\mu$ beam has a $\nu_e$ contamination at 1-2% level from the decay of the accompanying μ,

$$\mu^+ \to e^+ \nu_e \bar{\nu}_\mu. \qquad (51)$$

Both these problems are overcome in off-axis experiments as shown below. For this reason the K2K experiment has been succeeded by the off-axis T2K experiment, while the off-axis successor to MINOS (i.e. the NOvA experiment) will start operation next year. They are based on the simple result following from eq. (48), i.e.

$$\frac{dE_\nu}{dE_\pi} \cong (1 - \frac{m_\mu^2}{m_\pi^2}) \frac{1 - \gamma^2 \theta^2}{(1 + \gamma^2 \theta^2)^2}. \qquad (52)$$

This means that the neutrino energy becomes practically independent of the pion energy at θ = 1/γ = $m_\pi/E_\pi$, resulting in a quasi-monochromatic neutrino beam with

$$E_\nu \cong (1 - \frac{m_\mu^2}{m_\pi^2}) \frac{m_\pi}{2\theta} \cong \frac{30 MeV}{\theta}. \qquad (53)$$

Fig 4 shows the energy spectrum of the T2K neutrino beam for the on-axis (θ =0) along with several off-axis configurations [7].

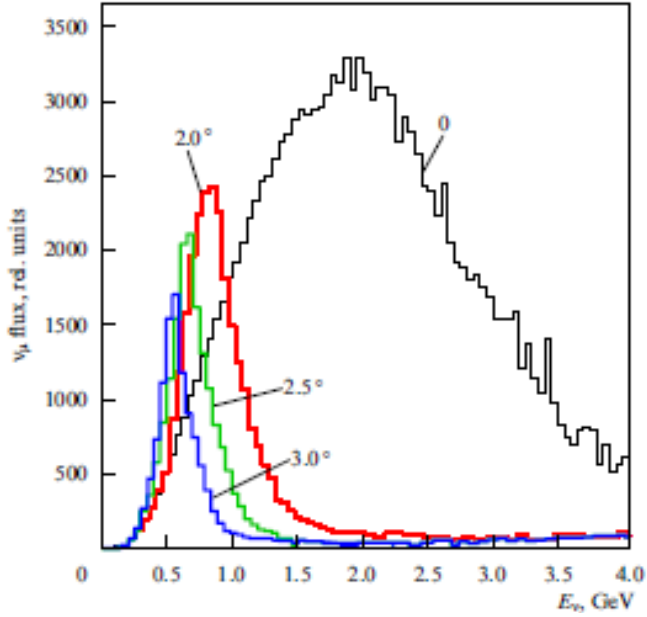

Fig 4. Energy spectra of the T2K neutrino beam at different angles relative to the proton beam axis : θ = 0 (on-axis), 2, 2.5 and 3 degrees. The T2K experiment uses the off-axis angle of θ = 2.5 degree [7].

The on-axis spectrum shows a broad peak at $E_\nu \approx 2$ GeV, corresponding to a broad peak at $E_\pi \approx 4$ GeV. This would correspond to an off-axis angle and energy

$$\theta = 0.035 = 2°, \quad E_\nu \approx 0.85 \text{ GeV}, \tag{54}$$

as shown in the figure. Note that the off-axis beam is quasi-monochromatic with very little spread in $E_\nu$, which effectively suppresses the neutral current background (50). The $\nu_e$ contamination from the secondary decay process (51) is also suppressed as it does not carry enough transverse momentum to reach the off-axis angle. Of course one has to pay the price in terms of a higher intensity of the proton beam to compensate for the decreased flux of the off-axis neutrino beam. Fig 4 also shows that one can tune the neutrino energy to still lower values by operating at a little larger off-axis angle of θ = 2.5 or 3 degree, in order to get closer to the maximal oscillation phase, $\Delta_{31} = 90°$. The T2K experiment operates at

$$\theta = 2.5° => E_\nu \approx 0.68 \text{ GeV}. \tag{55}$$

We shall now discuss the $\nu_\mu \to \nu_e$ appearance measurements by the MINOS, T2K and the forthcoming NOvA experiments.

**MINOS Experiment:** It is an on-axis experiment, designed for $\nu_\mu$ disappearance measurement, which it completed successfully. It uses the $\nu_\mu$ beam from Fermilab with a broad peak at $E_\nu = 3$ GeV. The 5.4 kt far detector is placed at a baseline length L = 725 km, while a 1 kt near detector is placed 1 km from the target to measure the $\nu_\mu$ flux and energy spectrum. Note that

$$L/E_\nu \approx 240 \text{ km/GeV} \Rightarrow |\Delta_{31}| \approx 42°, \qquad (56)$$

i.e. only half way to the oscillation maximum. Both the detectors are tracking sampling calorimeters with alternate layers of passive (steel) and active (plastic scintillator) materials embedded in a magnetic field. The scintillation light is collected with the help of wavelength shifting fibers and measured by PMT [23]. It detects the $\nu_e$ appearance signal via its charged current interaction in the iron layers

$$\nu_e \, N \to e^- X, \qquad (57)$$

producing EM shower [24]. The EM and hadron showers are distinguished by the shape of their energy distributions measured by the scintillator strips. However, there is a large uncertainty in the electron identification by this detector, resulting in a large neutral current background, apart from that of the $\nu_e$ contamination in the $\nu_\mu$ beam. The collaboration published results based on $8.2 \times 10^{20}$ POT (protons on target) data last year reporting 62 $\nu_e$ events [24]. With large estimated background of $50 \pm 8$ events it provided a $1.5\sigma$ signal for a nonzero $\theta_{13}$. The central value of this angle resulting from this signal is

$$\sin^2 2\theta_{13} = 0.04 \, (0.08) \text{ for +ve (-ve) } \Delta m_{31}^2 \qquad (58)$$

assuming $\delta = 0$. More recently the collaboration has updated their result with $10.7 \times 10^{20}$ POT data [25]. It reports 88 events against a background of $69 \pm 9$, which provides a $2\sigma$ signal for nonzero $\theta_{13}$. The central value is

$$\sin^2 2\theta_{13} = 0.06 \, (0.10) \text{ for +ve (-ve) } \Delta_{31}^2 \qquad (59)$$

assuming $\delta = 0$. The large dependence of this result on the sign of $\Delta m_{31}^2$ or equivalently $\Delta_{31}$ can be easily understood from eq. (27) using $A \approx \pm 0.3$ from (28) and the value of $|\Delta_{31}|$ from (56).

**T2K Experiment:** It is an off-axis neutrino experiment, which is optimized for $\nu_\mu \to \nu_e$ appearance. It uses the high intensity proton beam of the Japan Proton Accelerator Complex (J-PARC) with a beam power of 0.7 MW. It operates at an off-axis angle of $\theta = 2.5°$, corresponding to a quasi-monochromatic beam with peak $E_\nu \approx 0.68$ GeV, as we saw from eq. (55) and fig. 4. The $\nu_e$ contamination from the secondary decay process (51) is reduced to the level of 0.4%, while the neutral current background from (50) is also strongly suppressed. It uses a baseline length of L = 295 km for the far detector, corresponding to

$$L/E_\nu \approx 450 \text{ km/GeV} \Rightarrow |\Delta_{31}| \approx 80°, \qquad (60)$$

which is close to the maximal oscillation phase. The far detector is the famous SK detector, which is a 50 kt water Cherenkov detector surrounded by many thousands of PMT to measure the Cherenkov radiation energy [7, 26]. Two versatile multicomponent detectors are placed on-axis ($\theta = 0$) and off-axis ($\theta = 2.5°$) at 286 m from the target to measure the initial neutrino beam spectrum and continuously monitor its properties. They also provide accurate measurements of the differential cross-sections for all the charged and neutral current interactions, which is not possible with the SK detector. These measurements are used to estimate the backgrounds to the $\nu_\mu \to \nu_e$ appearance signal. The signal is detected in the SK detector via the charged current interaction, which is dominated at this beam energy by the quasi-elastic process

$$\nu_e (\nu_\mu) \, p \to e^- (\mu^-) \, n \qquad (61)$$

resulting in single Cherenkov ring events. The electron ring is distinguished from the muon ring by its diffused nature at ~ 1% level for such events. Selecting single Cherenkov ring events with electron like ring reduces the estimated background from the near detector data to the level of $\nu_\mu \to \nu_e$ appearance signal for $\sin^2 2\theta_{13} \approx 0.1$. With some further cuts on the event topology and the reconstructed neutrino energy reduces the latter to about 1/3$^{rd}$ of the signal size. The first result from the T2K experiment based on 1.5x10$^{20}$ POT data [26] reported 6 events against an estimated background of 1.5 ± 0.3 (syst), which provided a 2.5σ signal for nonzero $\theta_{13}$. The resulting central value of this angle was

$$\sin^2 2\theta_{13} = 0.11 \, (0.14) \text{ for +ve (-ve) } \Delta m_{31}^2 \qquad (62)$$

assuming $\delta = 0$. Recently the collaboration has presented results from 3x10$^{20}$ POT data [27] reporting 11 events against background of 3.2 ± 0.4 (syst), coming mainly from $\nu_e$ contamination (1.7) and neutral current (1.3) events. This constitutes a 3.2σ signal for nonzero $\theta_{13}$, with

$$\sin^2 2\theta_{13} = 0.094^{+0.053}_{-0.040} \left( 0.116^{+0.063}_{-0.049} \right) \text{ for +ve (-ve) } \Delta m_{31}^2 \qquad (63)$$

assuming $\delta = 0$. In either case there is a ±20% variation of the central value over the full cycle of δ. The relative insensitivity of the result to the mass hierarchy in this case is a reflection of the small baseline length or equivalently the small optimized beam energy of $E_\nu \approx 0.68$ GeV, corresponding to $A \approx ±6.8\%$. This is small enough for using the the 1$^{st}$ order formula (29). One can easily see from this formula a matter effect of $\approx ±10\%$ for the above values of A and $|\Delta_{31}|$. It accounts for the relative size of the two central values in eq. (63). Similarly for small A one can understand the ±20% variation of the central value over the full range of δ from the relative size of the first and second term in eq (19).

The T2K experiment plans to achieve a 26 fold increase in data to 7.8x10$^{21}$ POT over the next 5 years [27, 28]. By that time the $\sin^2 2\theta_{13}$ measurement from the reactor neutrino experiments would have reached an accuracy of ~ 5%. The projected $\nu_\mu \to \nu_e$ signal from this T2K data is shown over the full cycle of δ in fig. 5 along with the $\sin^2 2\theta_{13}$ measured in reactor neutrino experiments. As one sees from this figure, a comparison of the two results would be able to find a nonzero δ signaling CP violation at the 90% CL level over typically half the δ cycle.

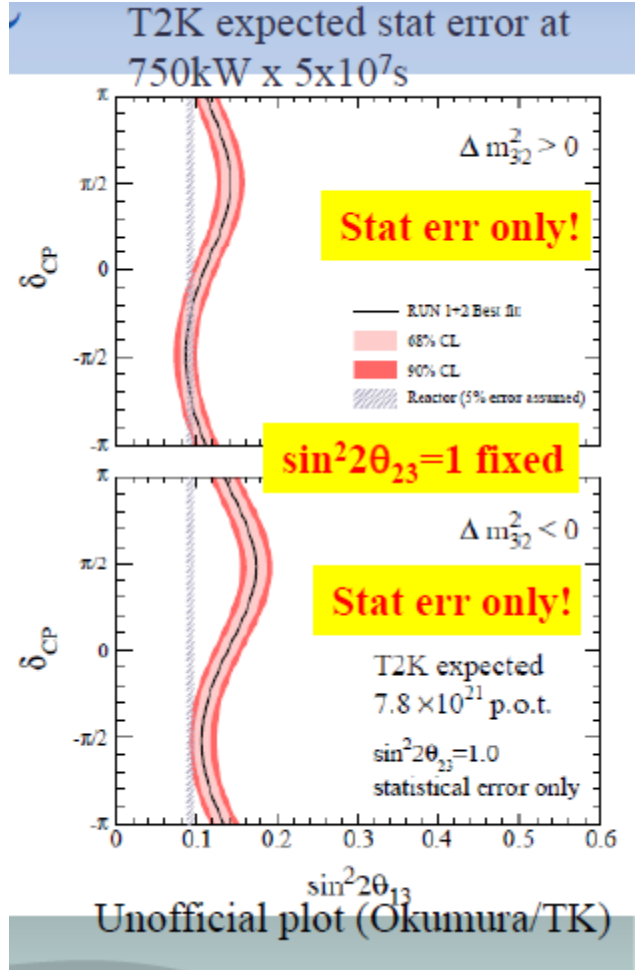

Fig 5. Projected $\nu_\mu \to \nu_e$ appearance signal from the full 7.8x10$^{21}$ POT data of the T2K experiment shown over the full cycle of δ for both normal and inverted mass hierarchies. The projected $\sin^2 2\theta_{13}$ band from the reactor measurement is also shown for comparison [28].

There are also plans to achieve large increase of data in future by increasing the intensity of the beam and/or the size of the detector to a ~ 1 Mt Hyper-Kamiokande water Cherenkov detector [29]. This will make it possible to select only a few regions of the δ parameter space depending on the mass hierarchy (sign of $\Delta m_{31}^2$). The mass hierarchy itself can be determined at 3σ level by the atmospheric neutrino data at the Hyper-Kamiokande detector, as discussed in the next section. Besides, there are proposals to extend the experiment by adding another far detector at Okinoshima, at a baseline length of 658 km and off-axis angle of 0.78° [28]. As one sees from (53), this would triple the beam energy to $E_\nu \approx 2$ GeV and the resulting sensitivity to the mass hierarchy by a similar factor. Thus one can combine the two far detector data to simultaneously determine the mass hierarchy and the value of δ.

**NOvA Experiment:** This is the off-axis sequel to the MINOS experiment, which will start operation from next year (2013). It is optimized for $\nu_\mu \to \nu_e$ appearance measurements using the high intensity proton beam from Fermilab [7, 30], with a beam power of 0.7 MW like J-PARC. It will operate at an off-axis angle of $\theta \approx 0.8°$, corresponding to a quasi-monochromatic beam with peak $E_\nu \approx 2$ GeV as per eq. (53). The far detector is located at a baseline length of L = 810 km, corresponding to

$$L/E_\nu \approx 405 \text{ km/GeV} \Rightarrow |\Delta_{31}| \approx 70°, \tag{64}$$

which is still fairly close to the maximal oscillation phase. The far and the near detectors are fully active segmented scintillation detectors of similar design, weighing about 14 kt and 0.3 kt respectively [31]. The main elements are long and narrow plastic cells filled with liquid scintillator, which are arranged horizontally and vertically in alternate layers. Each cell is connected by wavelength shifting fiber to a photodetector for measuring the scintillation energy. The $\nu_e$ and $\nu_\mu$ events are detected through the electron and muon produced via charged current interactions. The diffused profile of the electron track is well distinguished from the straight muon track in the scintillation detectors. The beam energy of $E_\nu \approx 2$ GeV corresponds to $A \approx \pm 0.2$ from (28), which is 3 times larger than that of the T2K experiment. Therefore one expects a matter effect of $\approx \pm 30\%$ for the $\nu_\mu \to \nu_e$ appearance probability of (29) for +ve (-ve) $\Delta m_{31}^2$. However, a modulation of similar size can also come from the variation of the CP violating phase $\delta$, which means that the two effects cannot be disentangled from the $\nu_\mu \to \nu_e$ data alone. Therefore the NOvA experiments plans to complete 3+3 years of $\nu_\mu \to \nu_e + \bar{\nu}_\mu \to \bar{\nu}_e$ appearance measurements. Recall that the $\bar{\nu}_\mu \to \bar{\nu}_e$ oscillation probability is obtained from that of $\nu_\mu \to \nu_e$ in eqs. (27) and (29) by changing $A \to -A$ and $\delta \to -\delta$. Thus one can in principle determine $\delta$ and the sign of A (i.e. $\Delta m_{31}^2$) by measuring both these oscillation probabilities. In practice, however, it will not be easy, as one sees from fig. 6 [31]. It shows the two predicted contours of $\nu_e$ and $\bar{\nu}_e$ appearance probabilities corresponding to the full cycle of $\delta$ for the two signs of $\Delta m_{31}^2$. The two contours appear to have only a small overlap. However, the typical $2\sigma$ error bars of these probabilities are expected to be $\sim \pm 0.015$ after 3+3 years run [31]. This corresponds to an effective overlap region of the two contours covering a little over half of each. This means that one can resolve the mass hierarchy only over a little less than half the $\delta$ cycle, centered around $\delta \approx \pi/2$ ($3\pi/2$) if the actual sign of $\Delta m_{31}^2$ is negative (positive). And it will not be possible to determine a nonzero value of $\delta$ at the $2\sigma$ level [31].

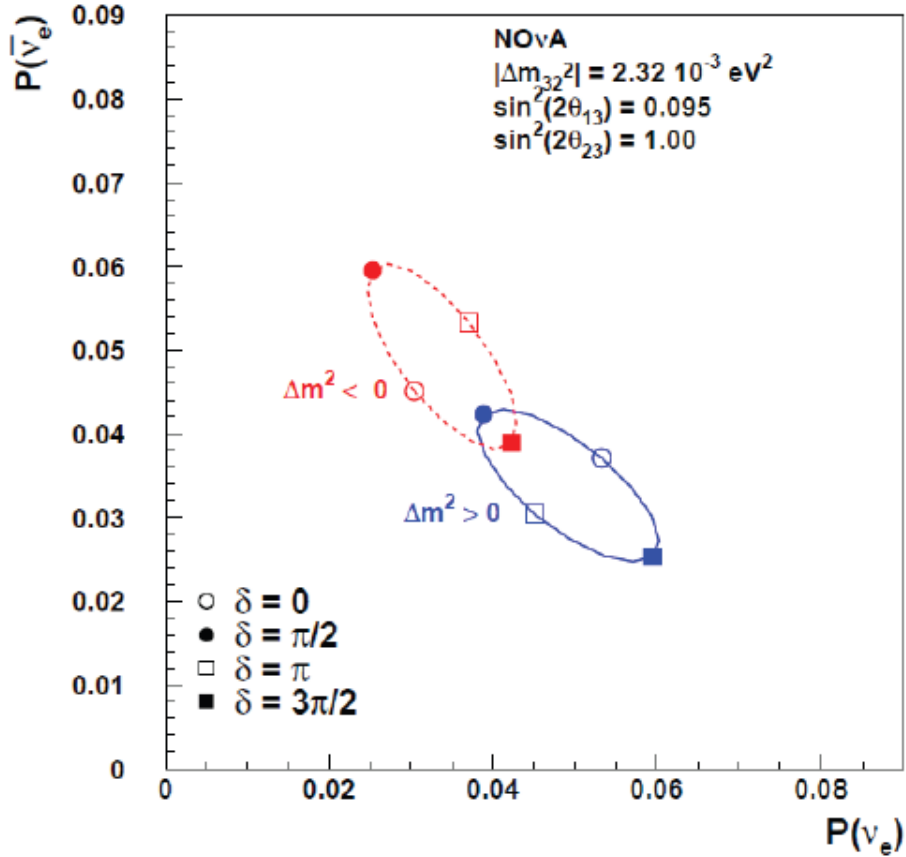

Fig 6. Predicted contours of the $\nu_e$ and $\bar{\nu}_e$ appearance probabilities covering the full cycle of δ for positive and negative signs of $\Delta m_{31}^2$. The typical 2σ error bars of these probabilities are expected to be ~ ±0.015 after 3+3 years of $\nu_e$ and $\bar{\nu}_e$ appearance measurements of NOvA [31].

More quantitative results can be found in ref [31]. It shows that combining the projected Nova and T2K data does not enhance the δ range of the 2σ resolution of mass hierarchy; but it provides a 1σ resolution over the entire cycle of δ. Moreover the combined data can observe a nonzero value of δ (signaling CP violation) at the 1.5σ (~ 90% CL) level over most of the δ cycle, although still not be able to reach the 2σ level [31]. Optimal ways of combining NOvA and T2K data have also been discussed in [32].

    Finally, there is a proposal for a new on-axis Long Baseline Neutrino Experiment (LBNE) using a new beam line from Fermilab with an initial beam power of 0.7 MW, which can be upgraded up to 2.2 MW [31]. It will have a 10 kt liquid Argonne Time Projection Chamber (TPC) as the far detector located at the Homestake mines, at a baseline length of 1300 km. It will be able to resolve the mass hierarchy at ≥ 2σ level over the entire δ cycle on its own, and at ≥ 4σ level in combination with NOvA and T2K data [31]. Moreover it will be able to determine a nonzero δ, signaling CP violation, over most of its parameter space (0.2π < |δ| < 0.8π) at ≥ 2σ level on its own, and at ≥ 3σ level in combination with the NOvA and T2K data [31]. As an alternative to the LBNE experiment there is a proposal to upgrade NOvA by installing a 30 kt liquid Argonne

detector at its far site. The combined data from this experiment along with those of the original NOvA and T2K experiments will be able to resolve the mass hierarchy at ≥ 2σ level over the full δ cycle and also observe a nonzero δ signal at ≥ 2σ level over the range 0.2π < |δ| < 0.8π [31]. In summary, we hope to determine the mass hierarchy (sign of $\Delta m_{31}^2$) and the CP violating phase with the data from T2K and NOvA experiments along with their proposed upgrades/extensions over the next decade or two. Thanks to the fairly sizeable value of the third mixing angle ($\sin^2 2\theta_{13} \approx 0.1$), it will be possible to achieve this with the conventional superbeam experiments in the foreseeable future instead of waiting for the beta beam or neutrino factory experiments. However, precision measurements of δ and other neutrino oscillation parameters and resolution of the remaining degeneracies will require these latter experiments of the more distant future.

## 6. Implications for Atmospheric Neutrino Experiments

Finally, let us discuss the prospect of resolving the mass hierarchy using the $\nu_\mu \to \nu_e$ or $\nu_e \to \nu_\mu$ measurements by the atmospheric neutrino experiments. The core-traversing atmospheric neutrinos travel a much larger distance through earth and encounter a larger terrestrial matter density than the LBL accelerator neutrinos. The constant matter density approximation and the resulting perturbative formula (27) is no longer valid in this case. Accurate analytic formulae for this case can be found e. g. in [33]. We shall summarize only the main results. The $\nu_\mu \to \nu_e$ and $\nu_e \to \nu_\mu$ appearance probabilities for core-traversing neutrinos have larger matter effect compared to the LBL accelerator neutrinos. And the magnitude of these appearance probabilities and their matter effects become fairly sizable for a sizable $\sin^2 2\theta_{13}$ ($\approx 0.1$). Moreover, they are fairly insensitive to the δ parameter unlike the LBL accelerator neutrino case. These advantages are offset, however, by two severe limitations of the atmospheric neutrino experiments compared to the LBL accelerator neutrino ones. Firstly, there is a huge background to the $\nu_\mu \to \nu_e$ ($\nu_e \to \nu_\mu$) appearance coming from the $\nu_e$ ($\nu_\mu$) survival probability, which is unsuppressed by any $\sin^2 2\theta_{13}$ factor. Secondly, the energy, direction and nature of the incoming neutrino have to be determined from the energy and direction of the final state particles along with the identification of the lepton (e/μ) and measurement of its charge. They make challenging demands on the detector performance of the atmospheric neutrino experiments. The $\nu_\mu$ appearance experiments offer accurate measurement of the muon charge (~ 95% purity) by magnetized iron tracking calorimeter, which is not possible for $\nu_e$ appearance experiments. On the other hand, the required accuracy of reconstructed neutrino energy and direction are much more demanding for the $\nu_\mu$ appearance experiments, because the $\nu_e \to \nu_\mu$ appearance probability varies rapidly with energy [33]. Therefore one requires a very high reconstructed neutrino energy and angular resolution of 5%, i.e.

$$\sigma_E / E_\nu = 0.05, \sigma_\theta = 5°, \qquad (65)$$

for a 2σ discrimination of the mass hierarchy with ~ 200 $\nu_\mu / \overline{\nu}_\mu$ events, while it will take an enormously larger data of ~ 6000 events to achieve this with a low resolution of 15% [33],

$$\sigma_E / E_\nu = 0.15, \sigma_\theta = 15°. \tag{66}$$

On the other hand, this 15% resolution would suffice for a 2σ discrimination of mass hierarchy with ~ 200 $\nu_e/\bar{\nu}_e$ events, if the experiment could separate the $\nu_e/\bar{\nu}_e$ events accurately. With a 80% purity of $\nu_e$ and $\bar{\nu}_e$ events the 2σ discrimination of mass hierarchy was estimated to require 1000-2000 $\nu_e/\bar{\nu}_e$ events [33].

Recently the SK collaboration has reported the $\nu_e/\bar{\nu}_e$ results based on their complete data, corresponding to 3900 days (240 kt-yr) exposure [34]. They can kinematically discriminate between the $\nu_e$ and $\bar{\nu}_e$ like events using the statistically larger fractional energy transfer (y) to hadrons for the neutrino events, since

$$d\sigma/dy_\nu = const., d\sigma/dy_{\bar{\nu}} \propto (1 - y_{\bar{\nu}})^2. \tag{67}$$

This implies a smaller energy fraction of the most energetic Cherenkov ring (e⁻), larger number of rings, larger p$_T$ and larger number of decay e$^\pm$ for the $\nu_e$ events. Combining this with the larger detection cross-section for $\nu_e$ events they are able to at least separate a sample of $\nu_e$ enriched events. For sin² 2θ$_{13}$ ≈ 0.1, one typically expects ~ 12% (5%) excess of core traversing $\nu_e$ events for normal (inverse) mass hierarchy from ν$_\mu$ → ν$_e$ appearance, while it is the other way around for $\bar{\nu}_e$ events [34]. However, with a data sample of over 2000 multi-GeV $\nu_e/\bar{\nu}_e$ events they are unable to resolve the mass hierarchy even at a decent fraction of 1σ level. Nor are they able to detect any statistically significant evidence of nonzero sin² 2θ$_{13}$, which does not require any $\nu_e/\bar{\nu}_e$ separation [34]. That means they are unable to find any statistically significant excess in the core-traversing $\nu_e/\bar{\nu}_e$ events, which would signal $\nu_e/\bar{\nu}_e$ appearance. The proposed 1 Mt scale Hyper-Kamiokande detector [29, 35] shall be able to resolve the mass hierarchy at the 3σ level from atmospheric $\nu_e/\bar{\nu}_e$ appearance experiment over a 10 year period irrespective of the δ parameter. Moreover, it will be able to determine δ as the far detector of the T2K experiment over this period [35].

It is pertinent to ask here whether the accurate muon charge measurement of a magnetized iron tracking calorimeter like MINOS or the proposed INO experiment [36, 37] can resolve the mass hierarchy with a modest data size of around ~ 200 multi-GeV $\nu_\mu/\bar{\nu}_\mu$ events. The MINOS experiment had observed atmospheric $\nu_\mu/\bar{\nu}_\mu$ events. But the size of the detector was too small to collect even this modest size data, since the typical number of such events is ~ 4/kt-yr. The proposed INO detector is scheduled to start operation in 2017 with a 50 kt magnetized iron tracking calorimeter. Assuming a fiducial volume of 30 kt, it will be able to accumulate 200-300 events in 2-3 years. It will be a tracking sampling calorimeter with alternate layers of passive and active materials like MINOS. The two main differences are that the active layers consist of Resistive Plate Chambers (RPC) instead of scintillators; and the passive layers consist of 5 cm thick iron plates which is twice the thickness of MINOS layers. The latter implies that the final state muon and hadron energies will be sampled by about half as many active layers in this case as in MINOS, resulting in a poorer energy resolution particularly for the hadron shower. The analysis of [33] has been updated in [38] for the INO detector with a less demanding high resolution criteria of 10%,

$$\sigma_E / E_\nu = 0.10, \sigma_\theta = 10^\circ, \tag{68}$$

along with the low resolution criteria of eq. (66). It found that a 50 kt INO detector with a high resolution power of 10% (68) can achieve 2σ discrimination of mass hierarchy on its own in ~10 years and in combination with the NOvA and T2K data in ~ 5 years for all values of δ. However, with a low resolution power of 15% (66) it can never achieve 2σ discrimination of mass hierarchy on its own; and it can only make marginal improvements to the reach of NOvA and T2K experiments. In the absence of the energy resolution estimate for the INO detector, one can get a rough idea from those of the MINOS detector [39], i.e.

$$\sigma_p / p = \sqrt{(0.06)^2 + (0.045 / p[GeV])^2}$$
$$\sigma_E / E \approx 0.55 / \sqrt{E[GeV]} \tag{69}$$

for the muon momentum and the hadron shower energy respectively. For a very rough estimate, we assume the average atmospheric neutrino energy for the experiment to be 6 GeV, shared equally between the muon and the hadrons as per the constant $d\sigma/dy_\nu$ of eq. (67), implying

$$\langle p_\mu \rangle \approx \langle E_{had} \rangle \approx 3 GeV. \tag{70}$$

Adding the resulting muon and hadron energy resolutions of (69) in quadrature gives

$$\sigma_{E_\nu} / E_\nu \approx 0.16, \tag{71}$$

which means the MINOS resolution of reconstructed $E_\nu$ is only ~ 16%. With coarser segmentation than MINOS the INO energy resolution would be poorer than this, in which case it would not be able to make any meaningful contribution to the resolution of the mass hierarchy. Let us conclude with the hope that the INO experiment will try to improve their energy and angular resolution at least to the level of MINOS, by considering e.g. a finer segmentation of their detector.

## 7. Summary

Till 2010 there were three unknown neutrino oscillation parameters – the third mixing angle $\theta_{13}$, the sign of the larger mass difference $\Delta m_{31}^2$, and the value of the CP violating phase δ. A number of indirect and direct experiments over the past two years have led to a consistent and by now fairly precise determination of $\theta_{13}$. Moreover, the value of this angle, $\sin^2 2\theta_{12} \approx 0.1$, is close to its earlier upper limit. This has promising implications for the determination of the two remaining unknown parameters through the present and proposed $\nu_\mu \to \nu_e$ appearance experiments in the foreseeable future. This pedagogical review starts with a brief introduction to the subject in section 1. Then section 2 discusses the relevant $\nu_\mu$ and $\nu_e$ disappearance and appearance formulae in the 3-neutrino oscillation formalism including the earth matter effect, which is responsible for determining the sign of $\Delta m_{31}^2$ (mass hierarchy). This is followed by a brief discussion of the first indication of a nonzero $\theta_{13}$ from the solar and KamLAND LBL reactor neutrino experiments in section 3. Then section 4 discusses direct determination $\theta_{13}$ this year (2012) by the three SBL reactor neutrino experiments of Double Chooz, RENO and especially Daya Bay. Section 5 discusses the $\nu_\mu \to \nu_e$ appearance measurement in the LBL accelerator neutrino experiments of MINOS, T2K and the forthcoming NOvA. The MINOS and especially T2K results provided valuable evidence for nonzero $\theta_{13}$ before the SBL reactor data; but the value of the resulting $\sin^2 2\theta_{13}$ was sensitive to the other two unknown parameters. Indeed the sensitivity of the LBL results to these remaining unknown parameters offers the promise of determining them from the forthcoming T2K and NOvA data along with those of their proposed upgrades/extensions using the precise measurement of $\sin^2 2\theta_{13}$ from reactor neutrino data as input. The T2K result is relatively insensitive to the mass hierarchy because of the relatively short baseline length. But one hopes to see a 90% CL ($\sim 1.5\sigma$) signal for the nonzero CP violating phase $\delta$ over about half of its parameter space after 5 years of T2K run. On the other hand, the proposed 3+3 years of $\nu_e + \bar{\nu}_e$ appearance measurements of the NOvA experiment offers a $2\sigma$ resolution of the mass hierarchy over nearly half the $\delta$ parameter space. Moreover both T2K and NOvA have proposed upgrades/extensions which offer definitive determination of the mass hierarchy along with the CP violating phase during the next decade or two. Thanks to the sizable value of $\sin^2 2\theta_{13}$, it will be possible to do this with the LBL accelerator neutrino experiments in the foreseeable future using the conventional superbeams. Section 6 discusses the prospect of determining the mass hierarchy with the $\nu_e$ and $\nu_\mu$ appearance measurements in atmospheric neutrino experiments using the large matter effect experienced by core-traversing neutrino. Unfortunately these experiments suffer from two serious limitations compared to the LBL accelerator neutrino experiments. Firstly the $\nu_e$ and $\nu_\mu$ appearance signals here have strong backgrounds from the respective survival probabilities, which are unsuppressed by any $\sin^2 2\theta_{13}$ factor. Secondly one has to reconstruct the energy, angle and type of the neutrino from those of the outgoing lepton and hadrons, which make challenging demands on the detector performance. Because of these reasons the complete data from the SK experiment with over 2000 multi-GeV $\nu_e / \bar{\nu}_e$ events has not been able to resolve the mass hierarchy even at a decent fraction of $1\sigma$ level. The proposed 1 Mt Hyper-Kamiokande detector offers a $3\sigma$ resolution of the mass hierarchy after 10 years run. The measurement of muon charge offers the advantage to atmospheric neutrino experiments with magnetized iron calorimeters like the proposed INO with relatively few $\nu_\mu / \bar{\nu}_\mu$ events. To exploit this advantage in practice, however, the experiment will need drastic improvement in its energy and angular resolutions of hadron shower over the envisaged ones. Let us conclude by mentioning two types of experiments not covered by this review. Firstly, the precision measurement of the neutrino oscillation parameters will require the LBL accelerator neutrino experiments of the more distant future using pure neutrino beams from neutrino factory or beta beam facilities. Moreover, there are aspects of neutrino physics, which are not amenable to oscillation experiments – notably determining the absolute scale of neutrino mass and its Majorana nature along with measuring the associated Majorana phases. Impressive progress has been made in addressing the former issue by the Katrin experiment and the latter ones by the various neutrinoless double beta decay experiments. However, definitive answers to these issues are likely to lie again in the more distant future.

## Acknowledgement

This review article developed from my discussions with the students participating in the NIUS (National Initiative in Undergraduate Science) camp of HBCSE (TIFR). The work was partly supported by the Indian National Science Academy through a senior scientist fellowship.